# Paint shop vehicle sequencing based on quantum computing considering color changeover and painting quality

**Jing Huang**

Department of Mechanical and Aerospace Engineering
University of Virginia
Charlottesville, VA 22904
jh3ex@virginia.edu

**Hua-Tzu Fan**

General Motors R&D
General Motors Corporation
Warren, MI 48090
charles.fan@gm.com

**Guoxian Xiao**

General Motors R&D
General Motors Corporation
Warren, MI 48090
guoxian.xiao@gm.com

**Qing Chang**[*]

Department of Mechanical and Aerospace Engineering
University of Virginia
Charlottesville, VA 22904
qc9nq@virginia.edu
*Corresponding author

**Abstract**

As customer demands become increasingly diverse, the colors and styles of vehicles offered by automotive companies have also grown substantially. It poses great challenges to design and management of automotive manufacturing system, among which is the proper sequencing of vehicles in everyday operation of the paint shop. With typically hundreds of vehicles in one shift, the paint shop sequencing problem is intractable in classical computing. In this paper, we propose to solve a general paint shop sequencing problem using state-of-the-art quantum computing algorithms. Most existing works are solely focused on reducing color changeover costs, i.e., costs incurred by different colors between consecutive vehicles. This work reveals that different sequencing of vehicles also significantly affects the quality performance of the painting process. We use a machine learning model pretrained on historical data to predict the probability of painting defect. The problem is formulated as a combinational optimization problem with two cost components, i.e., color changeover cost and repair cost. The problem is further converted to a quantum optimization problem and solved with Quantum Approximation Optimization Algorithm (QAOA). As a matter of fact, current quantum computers are still limited in accuracy and scalability. However, with a simplified case study, we demonstrate how the classic sequencing problem in paint shop can be formulated and solved using quantum computing and demonstrate the potential of quantum computing in solving real problems in manufacturing systems.

**Nomenclature**

| | | |
|---|---|---|
| | $v_i$ | The $i^{th}$ vehicle |
| | $c_i$ | The color of $i^{th}$ vehicle |
| | $s_i$ | The style of $i^{th}$ vehicle |
| | $\xi_i$ | Other information of $i^{th}$ vehicle |
| | $q_{ij}$ | Color changeover count switching from vehicle $v_j$ to $v_i$ |

| | |
|---|---|
| $r_{cc}$ | Color changeover cost |
| $\hat{f}(*)$ | Machine learning model predicting probability of repair |
| $p_{ij}$ | Probability of vehicle $v_i$ needing repair when pain $v_j$ |
| $r_{pr}$ | Cost of repair a vehicle with painting quality issue |
| $x_{it}$ | Decision variable indicating if vehicle $v_i$ is painted in position $t$ |
| $\alpha$ | Penalty term in unconstrainted optimization problem |
| $H_C$ | Cost Hamiltonian |
| $H_M$ | Mixer Hamiltonian |
| $\gamma, \beta$ | Parameters to be optimized in QAOA algorithm |
| $\sigma_j^x$ | Pauli-X operator |

# 1. Introduction

In the automotive industry, painting is one of the major manufacturing stages along with pressing, welding and general assembly [1]. A typical paint shop processes hundreds of vehicles of various colors and styles in a day. The sequencing of these vehicles is one of the most crucial problems in paint shop daily operation, as the paint shop sequencing is directly related to the productivity and profitability of the automotive manufacturing. There have been a lot of research works dedicated to the paint shop sequencing problem in the past decades. This paper makes unique contribution in its novel problem formulation by relating paint shop sequencing to not only color changeovers but also product quality, as well as its adoption of state-of-the-art quantum optimization algorithm to address the exploding problem scale in the real world.

Regarding the existing research works on paintshop sequencing, the problem objective varies from one to another, but the impacts of paint shop sequencing on product quality are not explicitly considered in most of the works. The most common objective of sequencing problem is to minimize the color changeovers, which is incurred when changing the paints between two consecutive vehicles of different colors [2,3]. When two consecutive vehicles are of different colors, the painting robots have to purge the previous color and load a new color, and thus causing material costs and potential production losses. There are other objectives are often included to formulate a multi-objective optimization problem, e.g., customer demand [4], environment awareness [5], utility work minimization [6], and makespan [7] etc. However, in our prior study it is found that the paint shop sequencing significantly impacts the quality performance of the painting process. Specifically, the probability of a painted vehicle having defects is dependent on not only its own property such as color but also that of its preceding vehicle. The quality consideration further complicates the paint shop sequencing problem, and hence novel problem formulation and optimization techniques are much needed to solve the problem in a more generalized setting.

Regarding the solution approach, many search techniques and heuristics-based optimization methods have been applied to different variations of paint shop sequencing problems. As the sequencing problem can be formulated as a graph traversal problem, search methods such as beam search [8] and local search [9] have been applied to sequencing problems. These search algorithms reduce the computational costs in exhaustive search while ensures the performance of the final solution to some extent. Another category of the optimization methods is the heuristics-based optimization algorithms including genetic algorithm [10,11], small-world optimization algorithm [12], and ant colony algorithm [4,13] etc. Last but not the least, reinforcement learning has recently been found to be a feasible approach to training intelligent agents to sequence vehicles in the paint shop. For example, Leng et al. [14] apply deep reinforcement learning to the sequencing problem with mix bank. They formulate the release-by-color procedure and the storage procedure as a sequential decision-making problem using the Markov Decision Process framework. However, when the problem scales up to hundreds of vehicles, which is a typical production target range for a paint shop within one shift or one day, those methods are very likely to suffer from long computing time and deficient solutions due to the curse of dimensionality in classical computing.

Quantum computing is expected to be a potential way to address the scalability issue in classical computing. Recent years have witnessed considerable investments in developing quantum computers by numerous technology companies and startups, and increased efforts in applying quantum computing to real-world problems from both academia and industry [15–17]. Quantum computing is a type of computation conducted based on the collective properties of quantum mechanics, including superposition, entanglement, and tunneling etc., to perform calculations. In classical computers, the basic element for computing is a bit, which represents either 0 or 1. In the contrast, quantum computers use a quantum bit, or qubit, to conduct computation. Taking advantage of the superposition, a qubit can represent 0 and 1 simultaneously, which means a quantum computer could be exponentially more powerful than a classical computer if other conditions are comparable. Therefore, quantum computing has long been expected to solve problems that are previously thought to be unsolvable using classical computers. Its potential applications include optimization, encryption, computational chemistry, financial modeling, and machine learning etc.

Therefore, there could be potential applications of quantum computing to paint shop sequencing problem, as it can also be formulated as an optimization problem. Some research works towards this direction have been reported. Streif et al. [18] applied quantum computing to a simplified version of paint shop sequencing problem called binary paint shop problem (BPSP). In BPSP, the task is to paint the vehicles in the order given by a random sequence, which means the problem does not involve shuffling and sequencing the vehicles. Each individual vehicle will appear twice and be painted with a different color each time The problem is reduced to choosing a color for a vehicle when it appears for the first time, also known as coloring, in a randomly generated sequence. The goal of BPSP is to find a coloring to minimize the number of color changes between adjacent vehicles in the sequence. BPSP is solved with QAOA as well. It successfully demonstrated the potential of quantum computing in solving paint shop sequencing problems. However, as the authors mentioned, in real-world industrial settings, coloring car bodies requires more than two color, and it remains as future work to find a suitable mapping for the generalized paint shop problem [18].

To this end, this paper is aimed at formulating a general paint shop sequencing problem and solve it using quantum computers. The contributions of this paper are: 1) we include the repair cost in the paint shop sequencing problem formulation. A machine learning model pretrained on historical data is used to predict the probability of quality defects. We will show that repair cost could dominate the sequencing cost in some scenarios so that color changeover might be insignificant in sequencing. 2) A general formulation of paint shop sequencing problem is proposed without setting any limits in number of colors or styles. 3) The problem is converted into the quantum paradigm. It is demonstrated that the problem can be effectively solved with state-of-the-art quantum optimization algorithms. We want to highlight that although the quantum hardware capacity in this stage is extremely limited, we provide a general procedure that can be immediately applied to real-world sequencing problem when commercial-scale quantum computers are available in the near future.

The remainder of this paper is organized as the follows: Section 2 introduces the general paint shop sequencing problem and associated costs. In Section 3, we formulate the problem as a combinational optimization problem, which is further converted to a quantum computing problem in Section 4. A simplified case study is shown in Section 5 and the conclusion is presented in Section 6.

# 2. Paint Shop Sequencing Problem Description

In this paper, we address a general paint shop sequencing problem without limitations in number of colors and styles. We consider $n$ vehicles that are awaiting processing in the paint shop, which could be production target in one shift or one day. Each of the vehicles is defined as

$$v_i = [c_i, s_i, \xi_i] \tag{1}$$

where $c_i$ denotes the color of vehicle $v_i$, e.g., red, white, or black etc., $s_i$ denotes the model and style of vehicle $v_i$, e.g., model A with sunroof, and $\xi_i$ denotes other information of the vehicle. These $n$ vehicles need to be painted sequentially in a painting booth in the paint shop, one at a time. The reason that an optimal painting sequence has to be found is that different sequences lead to different costs. In this paper, we consider the following two costs.

## 2.1. Color changeover cost

If two consecutive vehicles in the sequence are of different colors, the painting robots have to purge previous color and load a new one required by the subsequent vehicle. The additional purging operation would cause material and cycle time losses and is referred to as color changeover cost, denoted as $r_{cc}$ (per changeover). Given the colors of the $n$ vehicles, the changeover count on vehicle $v_i$ that is immediately painted after vehicle $v_j$ is

$$q_{ij} = \begin{cases} 1, & \text{if } c_i \neq c_j \\ 0, & \text{otherwise} \end{cases} \tag{2}$$

Therefore, the color changeover cost of switching from vehicle $v_j$ to vehicle $v_i$ is $q_{ij} r_{cc}$.

## 2.2. Painting repair cost

If a vehicle is found to have painting defects in quality examination, the vehicle has to be sent for a rework or repair. The cost is referred to as painting repair cost and denoted as $r_{pr}$ (per repair).

It is found in our preliminary study that painting quality is dependent on not only the property of the current vehicle, but also, surprisingly, the property of its preceding vehicle. This finding relates the painting quality performance to the paint shop sequencing.

Since the painting quality is not deterministic and unknown at the scheduling stage, we rely on a machine learning (ML) model pretrained on historical data to predict how much likely a vehicle would need repair given the sequence. The output of the machine learning model is the probability of quality issues that require a repair. The details of the machine learning model are out of the scope of this paper. Essentially, such a machine learning model $\hat{f}(*)$ maps vehicle subsequence to the probability of repair, i.e.,

$$p_{ij} = \hat{f}(\mathcal{X}|\theta) \tag{3}$$

where $p_{ij} \in [0,1]$ is the probability of vehicle $v_i$ needs repair if it is painted immediately after vehicle $v_j$, $\mathcal{X}$ denotes the necessary inputs to the machine learning model, and $\theta$ denotes machine learning model parameters trained through real dataset. Using the expectation, the painting repair cost of vehicle $v_i$ given its preceding vehicle $v_j$ is $p_{ij}r_{pr}$.

With the color changeover cost and vehicle repair cost defined in this section, the goal of the problem is to find an optimal sequence that minimizes the total cost given the production plan of the shift or the day. In the following section, we will formally put the problem in mathematical formulation.

# 3. Problem Formulation

In this section, we will formulate the paint shop sequencing problem as a combinational optimization problem by defining decision variables, constraints, and objective function.

## 3.1. Decision variables

We can view the sequencing process as putting each of the vehicle $v_i$, $i = 1,2,...,n$, in a particular position $t$, $t = 1,2,...,n$. Therefore, in this problem, the decision variable $x_{it}$ can be defined as:

$$x_{it} = \begin{cases} 1, & \text{if vechile } v_i \text{ is painted at position } t \\ 0, & \text{otherwise} \end{cases} \quad (4)$$

The sequence of $n$ vehicles is thus represented by an $n \times n$ matrix $\{x_{it}\}_{n \times n}$. For example, the following matrix represents the painting sequence of three vehicles.

$$x = \begin{bmatrix} 0 & 1 & 0 \\ 1 & 0 & 0 \\ 0 & 0 & 1 \end{bmatrix}$$

According to the matrix, vehicle $v_2$ will be painted first and followed by vehicle $v_1$. Subsequently, vehicle $v_3$ will be painted after vehicle $v_1$. This means that a general paint shop sequencing problem containing $n$ vehicles have $n^2$ decision variables. The exponential growth in problem scale is the major factor that hinders classical methods to find the optimal solution to the problem.

## 3.2. Problem constraints

It should be noted that an arbitrary matrix filled with 0's and 1's cannot always be interpreted as a painting sequence due to physical meaning of the matrix. There are two constraints that a matrix has to satisfy to make a valid sequence matrix. First, one vehicle should only be painted once, i.e.,

$$\sum_{t=1}^{n} x_{it} = 1, \forall i = 1,2,...,n \quad (5)$$

Second, the painting booth only paints one vehicle at a time, i.e.,

$$\sum_{i=1}^{n} x_{it} = 1, \forall t = 1,2,...,n \quad (6)$$

As for the format of the sequence matrix, the first constraint enforces that each row of the matrix contains one and only one element of 1, while the second enforces the same for each column.

## 3.3. Optimization problem formulation

The ultimate goal of the optimization problem is to find an optimal sequence represented by a matrix that gives the lowest total cost. As discussed in Section 2, the cost includes changeover cost and potential repair cost, both of which depends on current vehicle at position $t$ and the vehicle at its preceding position $t-1$. In summary, the paint shop sequencing problem can be formulated as a combinational optimization problem with the following objective function and constraints.

$$x^* = \arg\min_X \sum_{i=1,j=1}^{n} (q_{ij}r_{cc} + p_{ij}r_{pr}) \sum_{t=2}^{n} x_{j(t-1)}x_{it} \qquad (7)$$

$$s.t. \sum_{t=1}^{n} x_{it} = 1, \forall i = 1,2,\ldots,n$$

$$\sum_{i=1}^{n} x_{it} = 1, \forall t = 1,2,\ldots,n$$

The cost switching from vehicle $j$ to vehicle $i$ is $g_{ij}r_{cc} + p_{ij}r_{pr}$, since the machine learning outputs probability of repair, we use the expected repair cost, i.e., $p_{ij}r_{pr}$, in the objective function. In the objective function, the cost is accounted only when $x_{j(t-1)} = 1$ and $x_{it} = 1$, which means $v_j$ is painted at position $t-1$ and vehicle $v_i$ at position $t$.

# 4. Quantum Algorithm for Paint Shop Sequencing

In this section, we will solve the paint shop sequencing problem with a state-of-the-art quantum algorithm called Quantum Approximation Optimization Algorithm (QAOA). Since Farhi et al. [16] introduced QAOA in 2014, it has become a widely studied and adopted method for solving combinatorial optimization problems on Noisy Intermediate Scale Quantum (NISQ) devices, which means it is compatible with most of the existing universal quantum computers at current stage. We will first convert the optimization problem defined in previous section to unconstrainted form and then introduce the workflow of QAOA.

## 4.1. Objective function conversion

A convenient way to convert a constrained optimization problem to an unconstrained one is to add the constraints to the objective function as a penalty term.

$$C(x) = \sum_{i=1,j=1}^{n} (q_{ij}r_{cc} + p_{ij}r_{pr}) \sum_{t=2}^{n} x_{it}x_{j(t+1)} + \alpha \sum_{i=1}^{n} \left(\sum_{t=1}^{n} x_{it} - 1\right)^2 + \alpha \sum_{t=1}^{n} \left(\sum_{i=1}^{n} x_{it} - 1\right)^2 \qquad (8)$$

where $\alpha$ is a large constant, which dramatically increases the total cost if any of the constraints detailed in Eq. (5)-(6) are violated. Since this problem has $n^2$ decision variables, it corresponds to $n^2$ qubits in the quantum optimization algorithms. As of existing universal quantum computers, the maximum qubit number is extremely scarce. Therefore, this paper does not intend to solve the sequencing problem in the most practical scale. Rather, we aim to provide a general procedure to solve the sequencing problem in quantum computing.

## 4.2. QAOA to solve paint shop sequencing problem

In this subsection, we introduce the workflow of QAOA as applied to the general paint shop sequencing problem. We can map the objective function in Eq. (8) to the cost Hamiltonian.

$$H_C|x\rangle = C(x)|z\rangle \tag{9}$$

With this mapping, finding the optimal value of the objective function is equivalent to finding the extremal eigenvalues for the cost Hamiltonian. The cost Hamiltonian $H_C$ encodes the objective function $C$ and acts diagonally on the computational basis states of Hilbert space, i.e., $n^2$-qubit space in our sequencing problem for $n$ vehicles.

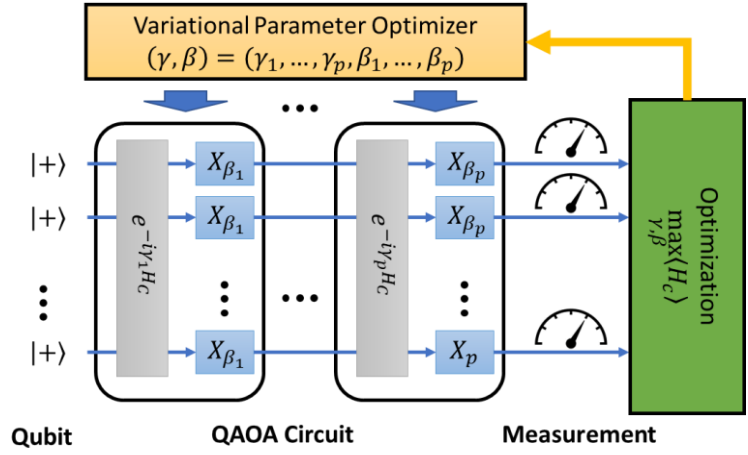

Figure 2. QAOA workflow for paint shop sequencing problem

Fig. 2 shows the workflow of a $p$-level QAOA. It starts from $n^2$ initial qubit. The cost operators in quantum gates are from the cost Hamiltonian:

$$U_C(\gamma) = e^{-i\gamma H_C} \tag{10}$$

where $\gamma$ is a parameter that will be optimized with a classical optimization algorithm. QAOA employs a mixer Hamiltonian $H_M$ that is defined as:

$$H_M = \sum_{j=1}^{n} \sigma_j^x \tag{11}$$

where $\sigma_j^x$ is the Pauli-X operator, which is the quantum equivalent of the NOT gate for classical computers with respect to the standard basis, i.e., $\sigma_j^x|0\rangle = |1\rangle$ and $\sigma_j^x|1\rangle = |0\rangle$.

$$U_M(\beta) = e^{-i\beta H_M} \tag{12}$$

Finally, the state of QAOA is obtained by applying the cost operator and mixer operator alternatively for $p$ times:

$$|\gamma,\beta\rangle = U_M(\beta_p)U_P(\gamma_p)\ldots U_M(\beta_1)U_P(\gamma_1)|s\rangle \tag{13}$$

where $p$ is the level of the QAOA. The optimal solution and value can be obtained by repeating the process of finding the optimal values of the variational parameters $\gamma$ and $\beta$. We will use a classical optimization algorithm to find such $\gamma$ and $\beta$ to get the minimum sequencing cost $C^*$.

$$C^* = \min_{\{\gamma,\beta\}}\langle\gamma_1,\beta_1|H_P|\gamma_p,\beta_p\rangle \tag{14}$$

QAOA shifts the variational parameter optimization steps to a classical computer, so that we find a way to make NISQ devices and classical computers work together given current limitations in quantum hardware [19].

## 5. Numerical Experiment

Currently, quantum computers are still confined to research purpose, as they have very low maximum qubit number and high error rate. The problem formulation and solution approach cannot be immediately applied to sequencing problem in the real world. However, in this section, we are able to conduct a simplified case study to demonstrate the proposed problem formulation and solution approach. The impact of paint quality cost on paint shop sequencing is also discussed.

### 5.1. Parameter setting

The case study is conducted on IBM quantum platform. Due to the qubit limit, we are only able to conduct a case study with three vehicles, which requires nine qubits, just fit the maximum qubits provided by the platform. Nevertheless, it suffices to validate the effectiveness of proposed quantum-based paintshop sequencing. In this case study, three vehicles of different colors and models are considered, i.e., $v_1 = \{\text{red}, \text{style } A\}$, $v_1 = \{\text{white}, \text{style } B\}$, $v_1 = \{\text{red}, \text{style } B\}$. Given the colors, the color changeover counts $q_{ij}$ are shown in Fig. 2. The matrix $\{q_{ij}\}_{n\times n}$ are symmetric due to the fact $q_{ij} = q_{ji}$.

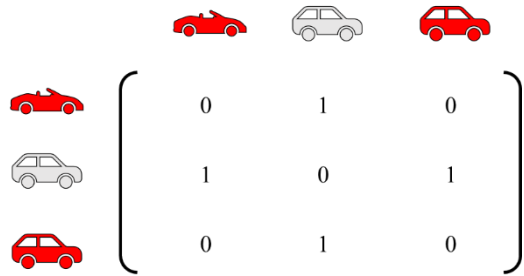

Figure 2. Color changeover count $\{q_{ij}\}_{n\times n}$

The repair probability $p_{ij}$ is the probability of vehicle $v_i$ needing repair given its preceding vehicle as $v_j$. The probability is outputted from a machine learning model. Fig. 3 shows the outputs from the machine learning model. Different from the color changeover counts, the repair probability matrix $\{p_{ij}\}_{n \times n}$ is asymmetric. Other case parameters include color changeover cost $r_{cc} = \$20$, repair cost $r_{pr} = \$100$. For QAOA, we choose $p$ to be 3 and use the classical optimizer by default in the IBM quantum platform.

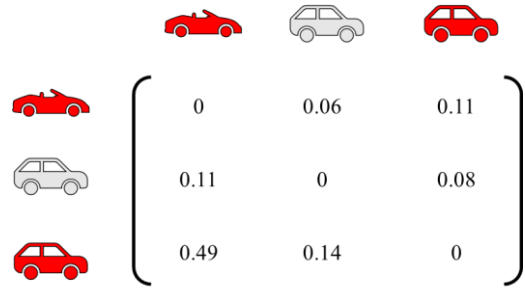

Figure 3. Repair probability $\{p_{ij}\}_{n \times n}$ outputted from machine learning model

## 5.2. QAOA results

The result from QAOA is shown in Fig. 4. Since there are only three vehicles, it is feasible to enumerate all the possible sequencings to obtain the optimal sequencing, which would be used to validate the result from QAOA. The QAOA arranges the sequence in a way that the total count of color changeover is 1, which is the least possible count in this case. The color changeover cost is thus $10. The potential repair cost is $11 for each switch. The final sequence is identical to the one we obtain using an exact method. Therefore, it turns out that the QAOA does deliver the optimal result, which demonstrates that the proposed paint shop sequencing problem formulation and quantum-based solution approach are correct.

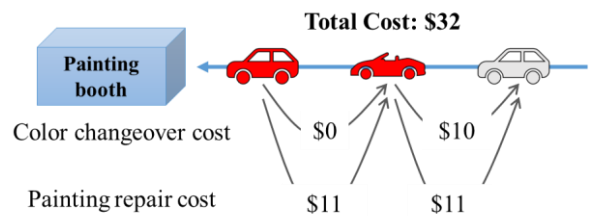

Figure 4. Optimal sequence produced by QAOA

## 5.3. Further analysis and discussion

In this paper, we incorporate the painting quality into paint shop sequencing problem. It is of interest to investigate how the repair cost would impact the sequence. Therefore, we change the paint repair cost from $100 per repair to $200 per repair and reconduct the experiment. As shown in Fig. 5, the optimal sequence now has two color changes, which means that the color change is increased to reduce the total paint repair cost. It demonstrates that the color change should not be considered as the sole indicator in a general paint shop sequencing problem since the impact of paint quality to the total cost is nontrivial.

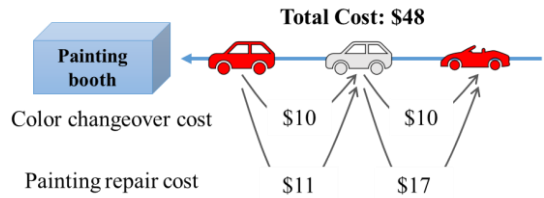

Figure 5. Optimal sequence with increased paint repair cost

There must exists a tipping point that the system has to increase color changeovers to reduce the paint repair cost. We can vary the paint repair cost and plot how total cost as well as each component changes with the paint repair cost as shown in Fig. 6.

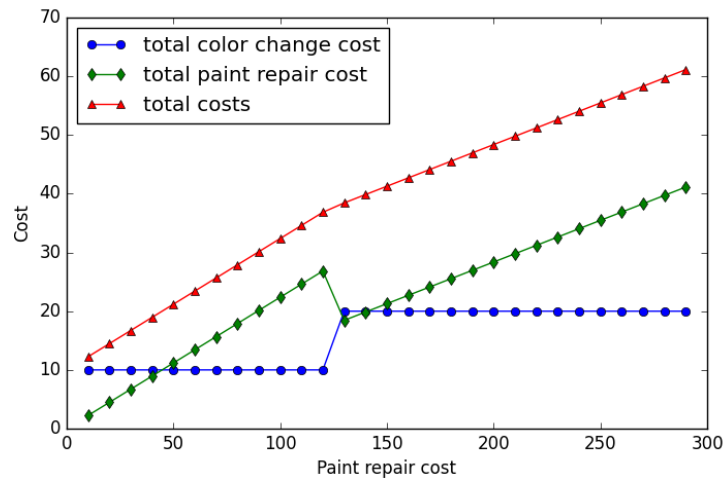

Figure 6. Optimal total costs changing with increased paint repair cost

It shows that when paint repair cost increases to $130, it becomes the dominant cost component in the problem and therefore color change cost is compromised. The insight into this observation is that one should not only focus on color changeovers when scheduling painting sequences. Sometimes, sequences with less color changeovers might lead to unstable painting quality and more repair work at the end of the line. On the other hand, in order to reduce color changes, one can either refine painting process to reduce repair probability or refine repair process to reduce cost per repair.

The unique properties in quantum mechanics, e.g., superposition, entangling, and tunneling etc., empower quantum computing to exceed the capability limits in classical computing. Thanks to the advantage of the superposition, the basic element in quantum computing, i.e., a qubit, can represent two states at the same time. A quantum computer could thus be exponentially more powerful than a classical computer, not only because the qubits represent more information than classical bits, but also due to the fact that the quantum computer can process the information more efficiently in an exponential order. As such, quantum computers have the potential to provide computational power on a scale that traditional computers cannot ever match. At some time point into the future when the quantum computers are fully commercialized, they could unsolvable problems using classical computers, only if the specific application problem is formulated appropriately. Due to the current limit of quantum platform, we only use this simplified case study to illustrate how to formulate a real manufacturing problem that can be solved using quantum computing, such as this paint shop sequencing problem.

# 6. Conclusions

In this paper, we discuss a general paint shop sequencing problem with arbitrarily many colors and styles. In addition to color changeover, we also consider the impacts of sequencing on paint shop quality, i.e., the sequencing can affect the probability of painting defect on one vehicle. The problem is formulated as a combinational optimization problem and then converted to a quantum optimization problem. Although due to the hardware limit, we are only able to demonstrate a simplified case study with three vehicles, the results suffice to validate that the problem formulation and solution approach are correct.

However, it should be noted that the quantum computing, especially universal quantum computers, is still in research stage and yet to be used to solve any large-scale problems in the real world. There are still a lot of research works to be done in terms of both hardware and algorithm. In the future, we will work with industrial partners to validate the approach and improve the algorithm in a larger scale with more qubits. The paint shop sequencing problem with multiple painting booths are also one of the potential future directions.

## Acknowledgements

This work is supported by National Science Foundation Grants 1853454.